\documentclass[aps,prb,twocolumn,groupedaddress,showpacs]{revtex4}
\usepackage{graphicx} 
\usepackage{psfrag} 
\usepackage{color}
\usepackage{amsbsy} 
\usepackage{amssymb} 
\usepackage{lineno}
\usepackage{chemarr}

\definecolor{purple}{rgb}{0.65, 0, 1}
\definecolor{orange}{rgb}{1,.5,0}

\def\ds{\displaystyle}
\def\Ec{\mathcal E}
\def\Nb{\mathbb N}
\def\Rb{\mathbb R}

\newcommand{\prob}[1]{\mathbb{P}\big{\{}#1\big{\}}}

\def\veps{\varepsilon}

\newcommand{\brak}[1]{\langle #1  \rangle}

\newcommand{\indic}[1]{\mathbf{1}_{#1}}

\begin{document}

\title[First Assembly Times]{First passage times in homogeneous
nucleation: dependence on the total number of particles}

\author{Romain Yvinec$^{1}$, Samuel Bernard$^{2,3}$, Erwan Hingant$^{4}$, Laurent Pujo-Menjouet$^{2,3}$}
\affiliation{$^1$ PRC INRA UMR85, CNRS UMR7247, Universit\'e Fran\c{c}ois Rabelais de Tours, IFCE, F-37380 Nouzilly}
\affiliation{$^{2}$ Universit\'e de Lyon, CNRS, Universit\'e  Lyon 1, Institut Camille Jordan UMR5208, 69622 Villeurbanne, France}
\affiliation{$^{3}$ INRIA Team Dracula, Inria Center Grenoble Rhône-Alpes, France}
\affiliation{$^{4}$ Departamento de Matem\'atica, Universidad Federal de Campina
Grande, PB, Brasil.}

\date{\today}

\begin{abstract}
\noindent
Motivated by nucleation and molecular aggregation in physical, chemical and biological settings, we present an extension to a thorough analysis of the stochastic self-assembly of a fixed number of identical particles in a finite volume. We study the statistic of times it requires for maximal clusters to be completed, starting from a pure-monomeric particle configuration. For finite volume, we extend previous analytical approaches to the case of \textit{arbitrary size-dependent} aggregation and fragmentation kinetic rates. For larger volume, we develop a scaling framework to study the behavior of the first assembly time as a function of the total quantity of particles.

\noindent We find that the mean time to first completion of a maximum-sized cluster may have surprisingly a very weak dependency on the total number of particles. We highlight how the higher statistic (variance, distribution) of the first passage time may still help to infer key parameters (such as the size of the maximum cluster) from data. And last but not least, we present a framework to quantify the formation of cluster of macroscopic size, whose formation is (asymptotically) very unlikely and occurs as a large deviation phenomenon from the mean-field limit. We argue that this framework is suitable to describe phase transition phenomena, as \textit{inherent infrequent stochastic processes}, in contrast to classical nucleation theory.
\end{abstract}

\maketitle

\section{Introduction} 
\label{S:intro}

\noindent
The self-assembly of macromolecules and particles into cluster is a fundamental process in many physical, chemical an biological systems.
Although particle nucleation and assembly have been studied for many decades \cite{BD,KUIPERS}, interest in this field has recently intensified due to
engineering, biotechnological and imaging advances at the nanoscale level \cite{WHITESIDES0, WHITESIDES1, GROS}. Applications range from material physics to cell physiology and virology (for a detailed list of examples see \cite{YOC} and references therein). Many of these applications involve a fixed ``maximum" cluster size -- of tens to hundreds of units -- at which the process is completed or beyond which the dynamics change \cite{OURJCP,Holcman2015}. One example include the rare self-assembly of mis-folded proteins into fibril aggregate in neurodegenerative diseases (Alzheimer, Parkinson, Prion...) \cite{SOTO,NOWAK}. Developing a stochastic self-assembly model focusing on the formation of a fixed ``maximum'' cluster size is thus important for our understanding of a large class of biological processes, and the quantification of the variability of the experimental data \cite{powers06,Xue08,Hingant,YVINEC,Knowles10}.

\noindent Theoretical models for self-assembly have typically described mean-field concentrations of clusters of all possible sizes using the well-studied mass-action, Becker-D\"{o}ring equations \cite{PENROSE,WATTIS,SMEREKA,OURPRE}.  While Master equations for the fully stochastic nucleation and growth problem have been derived, and initial analyses and simulations performed \cite{Marcus:1968, Lushnikov:1978, Kelly,BHATT,EBELING}, there has been relatively less work on the stochastic self-assembly problem. It has been recently shown that in
finite systems, where the maximum cluster size is capped, results from mass-action equations are inaccurate and that in this case a discrete
stochastic treatment is necessary \cite{DLC,YOC}. We consider here the Becker-D\"oring model (BD) defined by the following biochemical reactions
\begin{equation}\label{eq:chemreact_intro}
  \ds C_1 + C_i \ds \xrightleftharpoons[q_{i+1}]{p_{i}}  \ds C_{i+1} \,, \quad i\geq 1\,,
\end{equation}
where $C_i$ denotes the number (or concentration) of clusters of size $i$. Note that in the stochastic version (SBD), the state-space of the system is discrete and finite (see Fig.~\ref{FIG1}), given by all possible combinations that have a given fixed total number of particles (defined by $M$, given by the initial condition)
\begin{equation}\label{eq:space_stoch_E}
\Ec := \Big\{ (C_i)_{i \geq 1} \subset \Nb \, : \,\sum_{i\geq 1} i C_i = M \Big\}\,.
\end{equation}
\noindent The configuration $(C_i(t))_{i \geq 1}$ evolve in continuous time by discrete jumps according to Markovian description of the reactions \eqref{eq:chemreact_intro}. In our previous examination of first assembly time in this model \cite{YOC}, we found that a
striking finite-size effect arises in the limit of slow self-assembly. In particular, a \textit{faster} detachment rate can lead to a \textit{shorter} assembly time. This unexpected effect arise as the finite-size system may occupy some configurations that have been named ``traps'', where no free particle is available and the maximal-size cluster completion may occur only through first a detachment of a particle from a cluster. Discrepancies between the mean-field mass-action approach and the stochastic model were indeed most apparent in the strong binding limit.
\noindent In this paper, we will be interested in the generalization of earlier results\cite{YOC} on the distribution of the first
assembly times towards the completion of a full cluster, for arbitrary aggregation and fragmentation rates. Indeed, constant-size reaction rates was the main limitation of previous studies \cite{YOC}, as both physical and biological modeling require in many cases size-dependent attachment and detachment rates \cite{calvez10,Hingant}. Moreover, we will focus here on the dependency of the assembly times on the total initial number of monomers $M$. We will show how statistics of the first assembly time as a function of the total number of monomers $M$ may shed light on the biophysical properties of the appeared critical aggregates (size and size-dependence reaction rates). And we will highlight the discrepancies between the mean-field mass-action approach and the stochastic model \textit{even in the presence of a high number of monomers $M$ and its large limit}.



\noindent In the next section, we review the Becker-D\"{o}ring mass-action
equations for self-assembly and introduce the full stochastic problem. We derive the stochastic equations for the time-dependent cluster numbers, and introduce assembly times as first passage time problems. In Section III, we explore two simplified models for which the first assembly time can be solved analytically, and derive asymptotic expressions for the first assembly time in both the large number of monomer limit and large cluster size limit. Results from kinetic Monte-Carlo simulations (or Gillespie's algorithm) are presented in Section IV and compared with our analytical estimates. Finally, we discuss the implications of our results and propose further extensions in the Summary and Conclusions.

\section{Stochastic Becker-D\"{o}ring model, First assembly times definitions}

\noindent
The classic deterministic mass-action description for spontaneous, homogeneous self-assembly is the Becker-D\"{o}ring model \cite{BD} (BD), where the concentrations $c_{k}(t)$ of clusters of size $k$ obey an infinite (or truncated up to $k=N$) system of ordinary differential equation, given, for all $t\geq 0$, by
\begin{equation}\label{eq:detBD}
\left\lbrace 
\begin{array}{rcl}
\ds \frac{d}{dt}c_{1}(t) & \ds = & \ds -2j_1(t)-\sum_{k\geq 2} j_k(t)\,, \\
\ds \frac{d}{dt}c_{k}(t) & \ds = & \ds j_{k-1}(t)-j_k(t)\,,\quad k\geq 2\,, \\
\end{array}
\right.
\end{equation}
with 
\begin{equation}\label{eq:detBD_flux}
 \ds j_{k}(t)   =  p_{k}c_{1}(t)c_{k}(t)- q_{k+1} c_{k+1}(t)\,,\quad k\geq 1\,, \\
\end{equation}
and initial condition $\ds c_1(0) =  M$ and $\ds c_k(0)= 0$ for all $ \ds k\geq 2$. The rates $p_{k}$ and $q_k$ are respectively the monomer attachment and detachment rates to and from a cluster of size $k$. These rates are limited to sub-linear function of $k$, with bounded increments, in order to fulfill the standard well-posedness criteria \cite{BALL,Laurencot02}. It has been previously shown that such equations provide a poor approximation of the expected number of clusters when the total mass $M$ and the maximum cluster size $N$ are comparable in magnitude \cite{DLC}. Furthermore, such representations do not capture the randomness of the binding/unbinding events and cannot  describe the heterogeneity of cluster size distribution and time-dependent property such as first assembly times.\textit{ A stochastic treatment is thus necessary and is the subject of the remainder of this paper}.

\begin{figure}[h!]
\begin{center}
\includegraphics[width=3.4in]{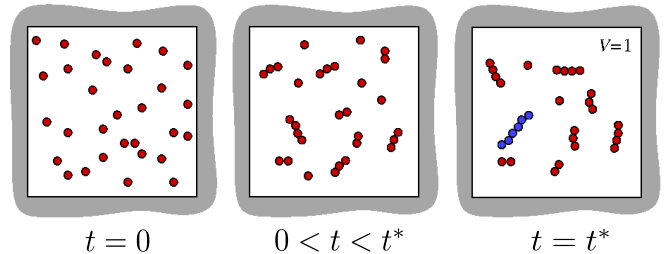}
\caption{Homogeneous self-assembly and growth in a closed unit volume
  initiated with $M=30$ free monomers. At a specific intermediate time
  $0 < t < t^*$
  in this depicted realization, there are six free monomers, four
  dimers, four trimers, and one cluster of size four. For each
  realization of this process, there will be a specific time $t^{*}$
  at which a maximum cluster of size $N=6$ in this example 
  is first formed
  (blue cluster). This time $t^*$ is a realization of the First Assembly Time (FAT, see definition in \eqref{eq:Nucl})}
\label{FIG1}
\end{center}
\end{figure}

\noindent Using a Markovian approach, we have previously derived \cite{YOC} the Forward Master equation to describe the probability that the system is in any given admissible configuration (see Eq.~\eqref{eq:space_stoch_E}) at times $t\geq0$. An equivalent formulation of this model is given by stochastic equations, driven by Poisson processes (as a continuous-time Markov Chain). This approach is natural to perform numerical simulations of sample path, and is more efficient to compute numerically first assembly times than the Master Equation approach. Moreover, this approach leads to natural comparison with deterministic systems when the total mass $M$ is large. Denoting by $Y_k^+$ (resp. $Y_k^-$) the standard Poisson process associated to the forward aggregation (resp. fragmentation) reaction of clusters of size $k$, the stochastic Becker-D\"{o}ring (SBD) equations for the time evolution of the number $C_{k}(t)$ of cluster of size $k$ are given for $t\geq0$ by (starting from a \textit{pure monomeric initial condition})
\begin{equation}\label{HOMOEQN0}
\left\lbrace 
\begin{array}{rcl}
\ds C_{1}(t) & \ds = & \ds M -2J_1(t)-\sum_{k\geq 2} J_k(t)\,,  \\
\ds C_{k}(t) & \ds = & \ds J_{k-1}(t)-J_k(t)\,,\quad k\geq 2\,, \\
\end{array}
\right.
\end{equation}
with 
\begin{equation}\label{HOMOEQN0_flux}
\begin{array}{rcl}
\ds J_{k}(t) & \ds = & \ds Y_k^+\Big{(}\int_0^t p_{k}C_{1}(s)(C_{k}(s)-\delta_k^1 )ds\Big{)} \\ & & \ds -Y_{k+1}^-\Big{(}\int_0^t q_{k+1} C_{k+1}(s)ds\Big{)}\,,\quad k\geq 1\,. \\
\end{array}
\end{equation}
where $\delta_k^1=1$ if $k=1$ and $\delta_k^1=0$ if $k>1$.
\noindent The first assembly time (FAT) for the stochastic discrete Becker-D\"oring is defined as a first passage time problem\cite{REDNERBOOK}
\begin{equation}\label{eq:Nucl}
 T^{N,M}_{1,0}:=\inf\{t\geq 0 : C_N(t) =1 \}\,.
\end{equation}
Hence the FAT is the first time to obtain a cluster of size $N$, starting with a pure single particle initial state, with $M$ particle (see Fig.~\ref{FIG1} for an example). To link with macroscopic definition of the nucleation time, we will also consider the generalized first assembly time (GFAT) problems
\begin{equation}\label{eq:Nucl_general}
 T^{N,M}_{\rho,h}:=\inf\{t\geq 0 : C_N(t) \geq \rho M^h \}\,,
\end{equation}
for given positive constant $\rho$ and $0\leq h\leq 1$. 
\noindent Here, we want to analyze the behavior of $T^{N,M}_{\rho,h}$ when $M\to \infty$, for fixed $N$, and when both $M,N\to \infty$.
This behavior will depend on scaling on the physical rates $p,q$, which may naturally depend on the total mass (or volume) of the system\cite{VANKAMP}. One way of computing the distribution of first assembly times is to consider the Backward Kolmogorov equation (BKE) describing the evolution of the configuration probabilities as a function of local changes from the initial configuration. The BKE Approach was taken in \cite{YOC}. It has the advantage to yields exact results for the full distribution of FAT, but it is strictly limited by the size of the system of equations, that grows exponentially with $M$. In this paper, we rely on exact calculation of simplified reduced models, limit theorems from Eq.~\eqref{HOMOEQN0} for large $M$ and $N$, and extensive numerical simulations of these equations.

\section{Results and Analysis}


\noindent Although the state-space~\eqref{eq:space_stoch_E} of the SBD model~\eqref{eq:chemreact_intro} is finite, the first passage problem defined by Eq.~\eqref{eq:Nucl_general} is in general a very difficult problem: see preliminary studies in \cite{YOC,Kelly,BHATT,EBELING}. There are two distinct simplifications that allow the problem to be analytically tractable. We present them here briefly in sections \textbf{A} and \textbf{B} (and generalize results from \cite{YOC}). Then we present two asymptotic results for large volume, $M\to\infty$, with either finite or infinite nucleus size in sections \textbf{C} and \textbf{D}, respectively. The strategy will be based on re-scaling procedure of the stochastic Eq.~\eqref{HOMOEQN0}. Numerical illustrations and results are postponed to the next section.

\subsection{Constant Monomer formulation}

The SBD model defined by Eq.~\eqref{HOMOEQN0} has the constant mass property
\begin{equation*}
 \sum_{k\geq 1} kC_k(t) \equiv  \sum_{k\geq 1} kC_k(0)=M\,, \quad t\geq 0\,.
\end{equation*}
In the original formulation of the BD model (sometimes used in the deterministic context\cite{WATTIS,BALL}), the total mass of the system is not preserved, but rather the quantity of free particles (think \textit{e.g.} of a source/sink that will keep instantaneously constant the quantity of available free particles). We will refer to this formulation as the constant monomer stochastic Becker-D\"oring model (CMSBD). We can represent it by the following reactions
\begin{equation}\label{eq:chemreact_constant}
\left\lbrace \begin{array}{lll}
  \ds \emptyset  & \ds \xrightleftharpoons[q_{2}]{p_{1} M(M-1) } & \ds C_{2} \,, \\
  \ds C_k & \ds \xrightleftharpoons[q_{k+1}]{p_{k}M} & \ds C_{k+1} \,, \quad i\geq 2\,.
 \end{array}
 \right.
\end{equation}
In the above formulation, $C_1(t)\equiv M$ is now a constant over time. Note that we expect such model to be close to the original SBD (for small times, up to the FAT) in the limit of large number of particle $M$. The main advantage of the constant monomer formulation is to be linear and hence analytically solvable. 
\noindent Indeed, it is known that for linear population model \cite{Kingman1969}, the number of individuals in each subclass of the population (starting with no individuals at time $0$), namely here $C_2(t),...C_N(t)...$, are independent Poisson random variable. Moreover, for this model~\eqref{eq:chemreact_constant}, the mean $c_2(t),...c_N(t)...$ are solution of the linear equation, given for all $t\geq 0$, by
\begin{equation} \label{eq:BD_C1cst_det}
\ds \frac{d}{dt} c_k(t) = j_{k-1}(t)-j_k(t)\,, \quad \forall k \geq 2\,,\\
\end{equation}
with
\begin{equation} \label{eq:BD_C1cst_det_flux}
\left\lbrace \begin{array}{rclr}
\ds j_1(t) &=& \ds p_1M(M-1)-q_2c_{2}(t)\,, \\
\ds j_k(t) &=& \ds p_{k}Mc_k(t)- q_{k+1}c_{k+1}(t)\, & \ds \forall k \geq 2\,, \\
\end{array}
\right.
\end{equation}
and initial condition $\ds c_k(0)= 0$ for all $ \ds k\geq 2$. Note that the last set of Eq.~\eqref{eq:BD_C1cst_det}-\eqref{eq:BD_C1cst_det_flux} is very close to the deterministic Becker-D\"oring model~\eqref{eq:detBD}-\eqref{eq:detBD_flux} taking $c_1\equiv M$. To calculate the FAT $T^{N,M}_{1,0}$ we use the survival function
\begin{multline*}
  S^{N,M}_{1,0}(t)  := \ds \prob{T^{N,M}_{1,0} > t} \\
  =  \prob{C_N(s)=0, s\leq t \mid C_k(0)=M\delta_{k=1}}\,.
\end{multline*}
Then, using an absorbing boundary condition at $k=N$ ($q_N = p_N =0$) together with the initial condition entail that $C_N(t)=0$ for some $t\geq0$ if and only if $C_N(s)=0$ for all $s\leq t$, so that
\[S^{N,M}_{1,0}(t) = \prob{C_N(t)=0 \mid C_k(0)=M\delta_{k=1}}\,.\]
Finally, since $C_N(t)$ is Poisson distributed (linear system) with mean $c_N(t)$, we have
\begin{equation}\label{eq:tlag_C1cst}
 S^{N,M}_{1,0}(t) = e^{-c_N(t)}\,.
\end{equation}
Equations~\eqref{eq:BD_C1cst_det}-\eqref{eq:BD_C1cst_det_flux} with the absorbing boundary at $k=N$ can rewritten as a linear system
\begin{equation} \label{eq:BD_linear}
\left \lbrace\begin{array}{rcl}
 \ds \dot{\textbf c }&=& \ds \textbf A \textbf c + \textbf B \,,\\
\ds \dot c_N(t) &=& \ds p_{N-1}Mc_{N-1}(t)\,,\\
\end{array}
\right.
\end{equation}
where $\textbf{c}=(c_2,c_3,\cdots,c_{n-1})^T$, $\textbf B = (p_1M(M-1),0,\cdots,0)^T$ and $\textbf{A}$ is a tridiagonal matrix with elements
 \begin{equation*}
\left \lbrace \begin{array}{rcl}
\ds a_{k,k}  & \ds = & \ds -q_{k+1}-p_{k+1}M\,,\\
\ds a_{k+1,k}  & \ds = & \ds p_{k+1}M\,,\\
\ds a_{k,k+1}  & \ds = & \ds q_{k+1}\,.\\
\end{array} \right.
\end{equation*}
%
The study of the linear system~\eqref{eq:BD_linear} has been performed both for the infinite dimensional case\cite{KREER} and for the truncated case\cite{DUNCAN}. In particular, it is shown that a similarity transformation
\begin{equation*}
\textbf{A}=\textbf{D} \textbf{S} \textbf{D}^{-1}  
\end{equation*}
with $D=diag(\sqrt{\tilde Q_kM^k})$ and $\tilde Q_k=\prod_{j=2}^k \frac{p_{j-1}}{q_j}$ leads to a matrix $\textbf{S}$ real symmetric tri-diagonal with non-zero elements on the sub and super-diagonal. Hence, classical linear algebra results shows that the eigenvalues of $\textbf{A}$ are real and distinct. Then, a general form of $c_N(t)$ is given by
\begin{equation*}
c_{N}(t)=p_{N-1}M\Big{[}\sum_{k=1}^{N-2}\alpha_kV^{(k)}_{N-2}\frac{e^{\lambda_k t}-1}{\lambda_k}-(\textbf A^{-1} \textbf B)_{N-2}t\Big{]}.
\end{equation*}
where $\lambda_k, V^k$ denotes the eigenelements of $\textbf A$ and $\alpha_k$ are determined by initial conditions. Analytical solutions are available\cite{YVINEC} for constant coefficient only (the matrix $\textbf A$ is in such case a Toeplitz matrix, with constant diagonal values, see Annex A.1). However, asymptotic expression are valid in the general cases. In particular, we have for small times $c_2(t)= p_1M(M-1)t + o(t)$ and
\begin{equation*}
 \dot c_k = p_{k-1}Mc_{k-1}+ o(t)\,,
\end{equation*}
so that, for $t\ll 1$,
\begin{equation*}
c_N(t)\approx_{t \ll 1} M^N\prod_{k=1}^{N-1}p_k  \frac{t^{N-1}}{(N-1)!}\,,
\end{equation*}
and Eq.~\eqref{eq:tlag_C1cst} is thus the survival function of a Weibull distribution, of shape parameter $k=N-1$ and scale parameter $\lambda=((N-1)!/(M^N\prod_{k=1}^{N-1}p_k))^{1/(N-1)}$. Hence, we get
 \begin{equation}\label{eq:meantime_constant}
 \brak{T^{N,M}_{1,0}} \approx_{M\to \infty}  \frac{\Gamma(1+1/(N-1))}{\Big{(}\prod_{k=1}^{N-1}p_k\Big{)}^{1/(N-1)}} \frac{((N-1)!)^{1/(N-1)}}{M^{1+1/(N-1)}}\,.
 \end{equation}
Variance formula for the Weibull distribution yields the asymptotic coefficient of variation (standard deviation over the mean)
  \begin{equation}\label{eq:cvtime_constant}
 cv_{T^{N,M}_{1,0}} \approx_{M\to \infty} \sqrt{ 2(N-1) \frac{\Gamma(2/(N-1)}{\Gamma(1/(N-1))^2} -1 }\,.
 \end{equation}
 Note in particular that the coefficient of variation do not vanish in large population, and that it is independent of the particular shape of the aggregation rate and depends only on the size of the maximal cluster $N$.

\noindent For the generalized first assembly time GFAT, similar time scale asymptotic on the Eq.~\eqref{eq:BD_linear} on the mean gives the following expression
 \begin{equation}\label{eq:meantime_constant_GFAT}
 \brak{ T^{N,M}_{\rho,h} } \approx_{M\to \infty} \frac{C(p,N)}{M}\frac{1}{M^{(1-h)/(N-1)}}\,,
 \end{equation}
 where $C(p,N)$ is a constant that depends only on $N$ and the aggregation rates $p_k,k\leq N$ (that can be made explicit if the full solution of Eq.~\eqref{eq:BD_linear} is known).
Those asymptotic expressions are illustrated in Annex (Fig.~A.1) where a perfect match is observed with numerical simulations.

\subsection{Single cluster model}

Another simplified model that can be analytically solved for the FAT problem is given by the assumption that a single cluster can be formed at a time\cite{YOC,penrose07}. We expect such model to be close to the original model when the fragmentation dominates, so that formation of many (large) cluster is unlikely. In such model, called the single-cluster stochastic Becker-D\"oring (SCSBD) model we may represent only the size of the single cluster, so that the state space is now one dimensional, being simply
\begin{equation*}
\Ec_1 :=  [1,\hdots,N]\,,
\end{equation*}
and the possible reactions are given by ($k$ denotes the size of the single cluster)
\begin{equation}\label{eq:reaction_oneagg}
\left \lbrace \begin{array}{rcl}
\ds k=1 & \ds \xrightleftharpoons[q_{2}]{p_{1} M(M-1) } & \ds k=2 \,,\\
\ds k  & \ds \xrightleftharpoons[q_{k+1}]{p_{k} (M-k) } & \ds k+1 \,, \quad k\geq 2\,.
\end{array}\right.
\end{equation}
\noindent In such a scenario, exact solution and classical First Passage Theory \cite{VANKAMP} gives (it is a one-dimensional discrete random walk)
 \begin{equation}\label{eq:sol_oneagg}
 \brak{T^{N,M}_{1,0}} = \sum_{i=1}^{N-1} \sum_{j=1}^i \frac{\prod_{k=j+1}^{i}q_k}{\prod_{k=j}^{i}p_k}\frac{1}{M^{\delta_j^1}\prod_{k=j}^{i}(M-k)}\,.
\end{equation}
\noindent In addition, general formula for variance and cumulative distribution function are available \cite{Gillespie80}. Those theoretical expressions are illustrated in Annex (Fig.~A.2) where a perfect match is observed with numerical simulations.

\noindent Asymptotic expressions of the mean assembly time is straightforwardly deduced from Eq.~\eqref{eq:sol_oneagg}. For instance, assume $q_k=\frac{\overline{q}_k}{\veps}$ and that $\veps\to 0$, the leading order of the mean assembly time is
\begin{equation*}
\brak{T^{N,M}_{1,0}} \approx_{\veps\to 0} \frac{1}{\veps^{N-2}}\frac{\prod_{k=2}^{N-1}\overline{q}_k}{\prod_{k=1}^{N-1}p_k \prod_{k=0}^{N-1}(M-k)}\,.
\end{equation*}
\noindent Also, one can show that in the asymptotic $\veps\to0$, for large fragmentation rate, the FAT $T^{N,M}_{1,0}$ is an exponential distribution \cite{YOC}.

\noindent Finally, for large $N$ and $M$, we can rescale the sum in Eq.~\eqref{eq:sol_oneagg} to obtain a suitable expression for the mean FAT when $N\to \infty$. Assume that the aggregation rates scale with $M$ so that $p_1= \overline{p}_1/M^2$, $p_k=\overline p_k / M$, $k\geq 2$. Then, let us introduce the rescaled size variable $x=k/N$, and define the rescaled kinetic rate
\begin{equation*} 
\begin{array}{rcl}
\ds \overline p(x) &=& \ds \sum_{k\geq 2} \overline p_k \indic{[k/N,(k+1)/N)}(x) \,,\\
\ds q(x) &=& \ds \sum_{k\geq 2} q_k \indic{[k/N,(k+1)/N)}(x) \,,
\end{array}
\end{equation*}
we have, for $N=\sqrt{M}\to \infty$ (see details in Annex, section A.2.2),
\begin{multline}\label{eq:sol_oneagg_Mtoinfty}
\brak{T^{\sqrt{M},M}_{1,0}} \approx_{M\to\infty} M\int_0^1\int_0^y \frac{e^{(y^2-z^2)/2}}{q(y)}\ldotp\\ \exp\Big{[}\sqrt{M}\int_z^y \ln \left(\frac{q(x)}{\overline p(x)}\right)dx\Big{]}dydz\,.
\end{multline}
In particular, when $q(x)>\overline p(x)$ on an interval of positive measure on $[0,1]$, the last expression~\eqref{eq:sol_oneagg_Mtoinfty} implies that the mean FAT to reach the \textit{macroscopic} size $x=1$ ($k=N=\sqrt{M}$) is exponentially large as $M\to\infty$. As an example, suppose that $q>\overline{p}$ are size-independent. Then, Eq.~\eqref{eq:sol_oneagg_Mtoinfty} becomes
\begin{equation*}
\brak{T^{\sqrt{M},M}_{1,0}} \approx_{M\to\infty} \frac{M}{q}\int_0^1\int_0^y e^{(y^2-z^2)/2} \left(\frac{q}{\overline p}\right)^{\sqrt{M}(y-z)}dydz\,.
\end{equation*}
Those theoretical expressions are illustrated in Annex (Fig.~A.3 and~A.4). Note that a different approach is to link the one-dimensional discrete random walk~\eqref{eq:reaction_oneagg} with a one-dimensional stochastic differential equation, and to use Large Deviation Theory to derive asymptotic FAT \cite{penrose07} (Annex, section A.2.3). This scaling approach and the link with a continuous size model when $N\to \infty$ will be taken on the full SBD model in section $\textbf{D}$.

\subsection{Large $M$, finite $N$}


\noindent In this section, we investigate the behavior of the SBD and its FAT when the total number of particles $M$ tends to infinity, while the size $N$ of the maximal cluster to reach stay finite. We distinguish two scenario, which yields distinct results. In the first one, the aggregation and fragmentation rate $p_k,q_k$ are taken independent of $M$. As the aggregation propensities increase with $M$, it is expected that the FAT decrease to $0$ as $M\to \infty$. The objective is to find valid asymptotic expression, and its dependence with respect to other parameters, like the maximal cluster size $N$ for instance. In the second case, the aggregation rate is scaled with the total number of particles, with $p_k=\frac{\overline p_k}{M}$. This scaling is motivated by classical system size expansion of chemical reaction networks\cite{VANKAMP}. As the total number of particles increases, the volume also increases so that the overall reaction propensities of the aggregation reactions stay constant. In such case, one expect to recover the deterministic first passage time of the classical deterministic BD model.

\noindent Let us now introduce our general rescaling strategy. The number of cluster of size $k$, given by $C_k$, are rescaled into
\begin{equation*}
 D_k^M(t)=\frac{C_k(t/M^\gamma)}{M}\,
\end{equation*}
with $\gamma$ a scaling coefficient to be chosen latter. Then, from Eq.~\eqref{HOMOEQN0}-\eqref{HOMOEQN0_flux}, we obtain, for any $t\geq 0$,
\begin{equation}\label{HOMOEQN0_rescaled}
\left\lbrace 
\begin{array}{rcl}
\ds D_{1}^M(t) & \ds = & \ds 1 -2J_1^M(t)-\sum_{k\geq 2} J_k^M(t)\,,  \\
\ds D_{k}^M(t) & \ds = & \ds J_{k-1}^M(t)-J_k^M(t)\,,\quad k\geq 2\,, \\
\end{array}
\right.
\end{equation}
with
\begin{multline}\label{HOMOEQN0_rescaled_flux}
 J_k^M(t)   =  \ds \frac{1}{M}Y_k^+\Big{(}\int_0^t M^{2-\gamma} p_{k}D_{1}^M(s)(D_{k}^M(s)-M^{-1}\delta_k^1 )ds\Big{)} \\ 
- \frac{1}{M}Y_{k+1}^-\Big{(}\int_0^t M^{1-\gamma} q_{k+1} D_{k+1}^M(s)ds\Big{)}\,,\quad k\geq 2\,.
\end{multline}
\noindent We recall a standard result of convergence of Poisson Processes (law of large numbers \cite{KURTZ}), that
\begin{equation*}
 \frac{1}{n}Y(nt)\rightharpoonup_{n\to\infty} t\,,
\end{equation*}
where $Y$ is a standard Poisson Process.

\subsubsection{No scaling of the aggregation rate}\label{ssec:noscale}
\noindent Using $\gamma=1$, and the standard law of large numbers applied to the Eq.~\eqref{HOMOEQN0_rescaled}-\eqref{HOMOEQN0_rescaled_flux},
we can show \cite{Jeon97} (see Annex, section 3)  that the sequence of stochastic processes $(D_k^M(t))$ converges, as $M\to \infty$, in a rigorous sense (in the trajectory space) to the solution of the irreversible aggregation deterministic model (BD with $q_k=0$), given, for all $t\geq 0$, by
\begin{equation} \label{eq:BD_det_aggirr}
\left\lbrace \begin{array}{rclr}
\ds \frac{d}{dt}d_1 &=& \ds -2j_1(t)-\sum_{k\geq 2}j_k(t)\,, &\\
\ds \frac{d}{dt}d_k &=& \ds j_{k-1}(t)-j_k(t)\,, & \ds \forall k \geq 2\,,
\end{array}
\right.
\end{equation}
with
\begin{equation} \label{eq:BD_det_aggirr_flux}
 \ds j_k(t) = p_{k}d_1d_k(t)\,, \forall k \geq 1\,,\\
\end{equation}
and initial condition $d_1(0)=1$ and $d_k(0)=0$, for all $k\geq 2$.
Intuitively, in the rescaled variable $D_k^M$, the aggregation process is much more favorable compared to the fragmentation because the number of free particles is very large. By definition of the GFAT Eq.~\eqref{eq:Nucl_general}, with $h=1$,
\begin{equation*}
 MT^{N,M}_{\rho,1} = \inf\{t\geq 0 : D^M_N(t) \geq \rho \}\,.
 \end{equation*}
Then, using the convergence of $(D_k^M(t))$, we obtain the following asymptotic behavior of the GFAT for $h=1$,
 \begin{equation}\label{eq:gfat_largeM_noscaling}
 \lim_{M\to \infty} MT^{N,M}_{\rho,1} = \inf\{t\geq 0 : d_N(t) \geq \rho \}\,.
 \end{equation}
The latter quantity is deterministic, and may be finite or infinite, according to the respective value of $p_k,N$ and $\rho$.

\begin{figure}[h!]
\centering
\includegraphics{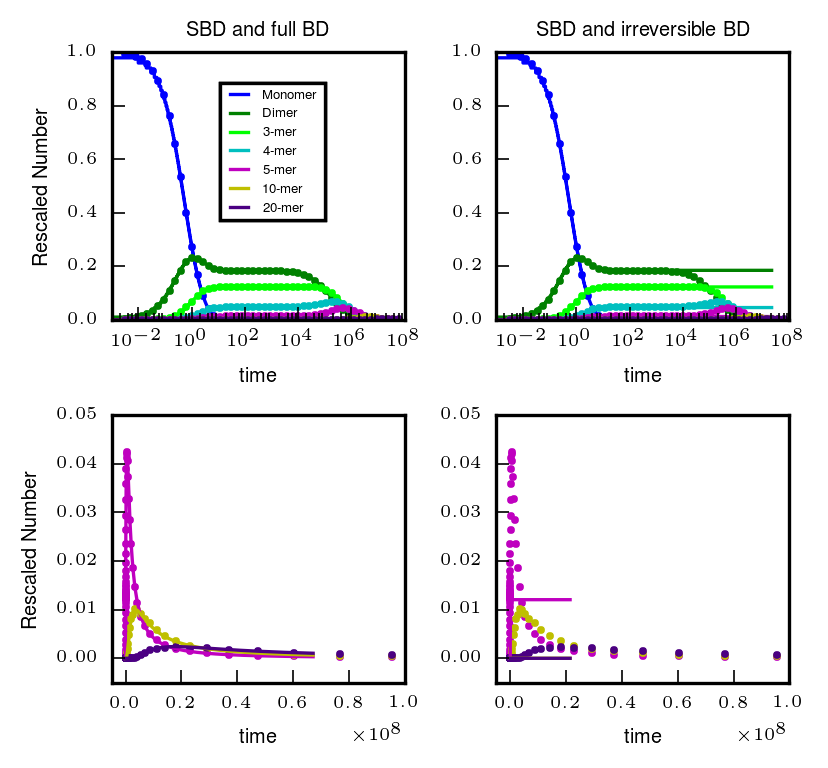}
\caption{SBD and BD Trajectories. For the SBD, we simulate the rescaled Eq.~\eqref{HOMOEQN0_rescaled}-\eqref{HOMOEQN0_rescaled_flux}, with $M=10^{5.5}$, and kinetic rates are $p_k\equiv 1$ for all $k\geq 1$ and $q_k\equiv 1$ for all $k\geq 2$. For the BD model, we simulate on the left columns the full BD Eq.~\eqref{eq:BD_det_aggirr}-\eqref{eq:detBD_favoragg_flux} and on the right columns the irreversible BD Eq.~\eqref{eq:BD_det_aggirr}-\eqref{eq:BD_det_aggirr_flux}. The rescaled SBD trajectories are plotted with filled circles, together with the corresponding BD trajectories in plain lines, for the monomer and $i$-cluster, $i=2,3,4,5,10,20$, according to the legend. The lower panel correspond to the same numerical simulation of the upper panel, with a zoom on the $y$-axis to improve the visaulization of the $i-cluster$ for $i=5,10,20$. It is immediate to see that the full BD Eq.~\eqref{eq:BD_det_aggirr}-\eqref{eq:detBD_favoragg_flux} agrees perfectly with the rescaled SBD Eq.~\eqref{HOMOEQN0_rescaled}-\eqref{HOMOEQN0_rescaled_flux} for all time, while the irreversible BD Eq.~\eqref{eq:BD_det_aggirr}-\eqref{eq:BD_det_aggirr_flux} matches only up to a time scale of order $M$.}
\label{fig:traj_det_sto_meta_mfinite} 
\end{figure}

\noindent The limit model~\eqref{eq:BD_det_aggirr}-\eqref{eq:BD_det_aggirr_flux} do not capture the FAT and the GFAT $T^{N,M}_{\rho,h}$ for $h<1$ (such event is reached for time $t=0^+$). However, as the initial number of monomers is large, we can use an intermediate approximation of the dynamic of the stochastic model, as a hybrid deterministic/stochastic model. First, note that pure coagulation BD model~\eqref{eq:BD_det_aggirr}-\eqref{eq:BD_det_aggirr_flux} have been extensively studied \cite{KRAP}, where exact time-dependent solution for $p_k=pk$ are given, and time asymptotic behavior are given for power law coefficient $p_k=pk^\lambda$, $0\leq \lambda \leq 1$. We restrict the following discussion to the constant rate case, $\lambda=0$, for simplicity (results are analogous in the power law case). In such case, the stationary state of the pure coagulation BD model\cite{KRAP,WATTIS}~\eqref{eq:BD_det_aggirr}-\eqref{eq:BD_det_aggirr_flux} is $d_1^*=0$ and
\begin{equation}\label{eq:meta_BDirr}
d_k^*=\frac{k-1}{e k!}\,,\quad k\geq 2\,.
\end{equation}
Although the rescaled threshold $\rho M^{h-1}$ will be reached by $d_N$ (and hence by $D_N^M$) for any $\rho$ and $h<1$ for large enough $M$ (as $d_N^*>0$), one can already see that for ``intermediate'' $M$, we may have $Md_N^* \ll 1$, so that the threshold may not have been reached while the free particle have vanished ($d_1^*=0$). In such case, it is necessary to take into account the small but crucial contribution of the aggregate shortening. For that, let us consider as a further approximation of Eq.~\eqref{HOMOEQN0_rescaled}-\eqref{HOMOEQN0_rescaled_flux} the following deterministic model (still with constant rate coefficients, to simplify the following), given, for all $t\geq 0$, by Eq.~\eqref{eq:BD_det_aggirr} and flux definition
\begin{equation}\label{eq:detBD_favoragg_flux}
\ds j_{k}(t) =  pd_{1}(t)d_{k}(t)- \frac{1}{M} q d_{k+1}(t)\,,\quad k\geq 1\,, \\
\end{equation}
where $1/M$ is seen as a small parameter. To obtain results for the quantity $T^{N,M}_{\rho,h}$ for any $h<1$, we need to study the time-dependent properties of the favorable aggregation limit $M\to \infty$ of the deterministic BD model Eq.~\eqref{eq:BD_det_aggirr}-\eqref{eq:detBD_favoragg_flux}. The following discussion is illustrated with Fig.~\ref{fig:traj_det_sto_meta_mfinite} (see also in Annex, Fig.~A.7,~A.8). For constant rate $p,q$, it is known \cite{WATTIS} that under favorable aggregation limit $q/M \ll p$, the deterministic BD model Eq.~\eqref{eq:BD_det_aggirr}-\eqref{eq:detBD_favoragg_flux} exhibits the following successive periods:\\
- Firstly, the model behaves as the irreversible aggregation BD model, Eq.~\eqref{eq:BD_det_aggirr}-\eqref{eq:BD_det_aggirr_flux}, during a time-scale of order $e \log(M)$, until monomer concentration $d_1(t)$ becomes small;\\
- Secondly, when the monomer concentration $d_1$ is of order $1/M$, there is a metastable period in which each concentration species  of size $k\geq 2$ are nearly constant, equal to $d_k^*$, the equilibrium state Eq.~\eqref{eq:meta_BDirr} of the irreversible aggregation BD model, Eq.~\eqref{eq:BD_det_aggirr}-\eqref{eq:BD_det_aggirr_flux}. The concentration $d_k(t)$ stays roughly constant to the values $d_k^*$, distinct from the steady-state values of the  full BD Eq.~\eqref{eq:BD_det_aggirr}-\eqref{eq:detBD_favoragg_flux}, until the next time scale;\\
- Thirdly, at a time scale of order $M$ (which is the time scale of aggregate shortening), larger aggregates are created within a process akin to diffusion in the size $k$-space (slow redistribution of aggregate sizes);\\
- Finally, every concentration species $d_k$ reaches the classical steady-state value of the  full BD Eq.~\eqref{eq:BD_det_aggirr}-\eqref{eq:detBD_favoragg_flux} within a time scale of order $M^2$. Steady-state values $\overline d_k$ are given by
\begin{equation*}
\overline d_k = \left( \frac{pM}{q}\right)^{k-1}\overline d_1^k\,,\quad k\geq 2\,,
\end{equation*}
where $\overline d_1$ is determined by the mass conservation property. Such values can be approximated by $\overline d_1 \approx 1/M$ and $\overline d_k\approx 1/M (1-1/\sqrt{M})^{k-1}$. 

\noindent To approximate the GFAT $T^{N,M}_{\rho,h}$, we need to know in which of these periods the event $\{C_N(t)\geq \rho M^h\}$ is reached. This mostly depends on the critical size $N$ of the nucleus as follows. If $M$ is large enough, then the metastable state is large enough, \textit{i.e.} $c_N^*=Md_N^*>\rho M^h$, and the cluster number $C_N(t)$ will reach the threshold during the pure-aggregation time-scale ($\log(M)/M$ in the original time scale), and the GFAT $T^{N,M}_{\rho,h}$ is found (see numerical section) to behave as the linear CMSBD model~\eqref{eq:chemreact_constant} with $C_1=M$. 

\noindent In the opposite case, for intermediate $M$ and large enough $N$ such that $c_N^*\ll \rho M^h$, we expect $C_N(t)$ to reach the threshold after the metastable period (of order $1$ in the original time scale). As the initial pure-aggregation phase is short ($\log(M)/M$), we can neglect it, and use the metastable values $c_k^*$ as initial values for a linear CMSBD model~\eqref{eq:chemreact_constant} where the monomer number is now equal to $C_1\equiv c_1^*$ (see Annex, section 3.1 and Fig.~A.10) given by\cite{WATTIS}
\begin{equation*}
 c_1^*=q\frac{c_2^*+\sum_{k=2}^{N-1}c_k^*}{p\sum_{k=2}^{N-1}c_k^*}=\frac{3}{2}\frac{q}{p}\,.
\end{equation*}
Hence $c_1^*$ is independent of the initial number of monomers $M$ and is of order $q/p$. Thus the GFAT depends on $M$ only through the initial condition $c_k^*$, $k\geq 2$, and is found to be (see numerical results) almost independent of $M$ on several order of magnitude for $N\geq 15$. Finally, note that there is always a (small) probability that the threshold is reached before the metastable period, which is responsible for a bimodal behavior of $T^{N,M}_{\rho,h}$  (see numerical results). For values of $d_k^*$ and a summary of the different cases, see Tables \ref{table:pi} and \ref{table:nucleationtime}.

\noindent Finally, performing a second-order approximation of Eq.~\eqref{HOMOEQN0_rescaled}-\eqref{HOMOEQN0_rescaled_flux}, we obtain a system of stochastic differential equation with variance of order $\sqrt{1/M}$ (see details in Annex, section A.3.2), and the GFAT can be computed by
\begin{small}
\begin{equation*}\label{eq:Nucl_sde}
 MT^{N,M}_{\rho,h} \approx \inf\{t\geq 0 : \tilde D_N^M(t) \geq \rho M^{h-1} \}\,,
\end{equation*}
\end{small}
\noindent where $(\tilde D_k^M)$ denotes the solution of the second order stochastic differential equations. Such approach unfortunately do not provide analytical hints on the behavior of the GFAT, but provide a convenient tool to compute numerically an approximation of the GFAT for very large $M$, where the exact MC simulations slow down (see numerical results).

\subsubsection{``System-size expansion'' scaling}\label{ssec:system-sizeexpansion}

The above asymptotic approximation have been performed assuming that the reaction rates are independent of $M$. However, the limit $M\to\infty$ may be understood as a system-size expansion, in which case reaction rates must be scaled with the system size according to their respective order. In particular, it is classical \cite{VANKAMP} that first-order reaction rates are independent of the system size, and second-order reaction rates are inversely proportional to the system size. Thus we are led to use $p_k=\frac{\overline p_k}{M}$. With $\gamma=0$, the re-scaled variable $D_k^M(t)=C_k(t)/M$ converges now to the  BD system given, for all $t\geq 0$, by Eq.~\eqref{eq:BD_det_aggirr} and flux definition
\begin{equation}\label{eq:detBD_limit_flux}
\ds j_{k}(t) =  \overline p_{k}d_1d_k(t)-q_kd_{k+1}(t)\,,\quad k\geq 1\,. \\
\end{equation}
\noindent As before, using the convergence of $(D_k^M(t))$, we obtain the following asymptotic behavior of the GFAT for $h=1$,
 \begin{equation*}
 \lim_{M\to\infty} T^{N,M}_{\rho,1} = \inf\{t\geq 0 : d_N(t) \geq \rho \}\,.
 \end{equation*}
Once again, the latter quantity is deterministic, and may be finite or infinite, according to the respective value of $q_k,\overline p_k,N$ and $\rho$. 
The GFAT $T^{N,M}_{\rho,h}$ with $h<1$ behaves asymptotically as the GFAT of the linear CMSBD model~\eqref{eq:chemreact_constant} with $C_1\equiv M$ and $p_k=\frac{\overline p_k}{M}$. Thus,
 \begin{equation}\label{eq:fpt_scaling_pk}
 T^{N,M}_{\rho,h} \approx_{M\to \infty} C(\overline p,N)\frac{1}{M^{(1-h)/(N-1)}}\,,
 \end{equation}
 where $C(\overline p,n)$ is a constant that depends only on $N$ and $\overline p(k),k\leq N$. Second-order approximation may also be derived using a central limit theorem for $D_k^M$ (see Annex, section A.3.2).

\subsection{Large $M$ and Large $N$} 
 

\noindent In this section, we investigate the behavior of the SBD and its FAT when the size $N$ of the maximal cluster is large, and scales with the total number of particles $M$. As in section \textbf{B}, we will then naturally use the rescale size variable $x=k/N$. We distinguish again two scenario, which yields distinct results. In the first one, the aggregation and fragmentation rates $p_k,q_k$ are taken independent of $M$.  In the second one, the aggregation rates are scaled with the total number of particles, with $p_k=\frac{\overline p_k}{M}$. In both cases, a rescaling of the solution is found to be solution of a deterministic continuous size model, namely the Lifshitz-Slyozov model (LS). Indeed, we have detailed in a companion paper \cite{DHY} how to choose a proper scaling and how to derive the limit equation for that rescaled solution. We show here the consistency of this scaling with the behavior of the GFAT.

\noindent We will restrict for simplicity here to the case $N=\sqrt{M}$. In that case, we define the rescaled measure on $\Rb^+$,

\begin{equation}\label{eq:rescaled_measure}
 \mu^M(t,dx)= \sum_{k\geq 2}\frac{C_k(t/M^\gamma)}{\sqrt{M}}\delta_{k/\sqrt{M}}(dx)\,,
\end{equation}
and $C_1^M(t)=C_1(t/M^\gamma)/M$, where $\delta_x(\cdot)$ is the Dirac measure at $x$. The GFAT $T^{\sqrt{M},M}_{\rho,h}$ involves a larger and larger maximal size $\sqrt{M}$, which is rescaled to the macroscopic size $x=1$ by the definition of the measure $\mu^M$ in Eq.~\eqref{eq:rescaled_measure}. We also need to define accordingly macroscopic aggregation and fragmentation rates, using
\begin{equation} \label{eq:macro_rate}
\left\lbrace \begin{array}{rcl}
\ds p^M(x) &=& \ds \sum_{k\geq 2}\overline p_k \indic{[k/\sqrt{M},(k+1)/\sqrt{M})}(x) \,,\\
\ds q^M(x) &=& \ds \sum_{k\geq 2} q_k \indic{[k/\sqrt{M},(k+1)/\sqrt{M})}(x) \,,
\end{array}
\right.
\end{equation}
where $\indic{I}(\cdot)$ is the characteristic function that equals $1$ in $I$ and $0$ outside.

\subsubsection{$N\to\infty$, No scaling of the aggregation rate}\label{ssec:Ngd_norescale}

Using $\gamma=1/2$, and a rescaled nucleation rate\footnote{The fact that the first aggregation rate need to be rescaled differently from the other aggregation rate comes from the special role played by the monomer in the BD and LS model. For a detailed discussion on the modelling point of view, see \cite{ColletGoudon,Collet}} $p_1=\frac{\overline p_1}{M}$, we have shown in \cite{DHY} that the that the measure $\mu^M$ satisfies
\begin{equation*}
 \lim_{M\to \infty} \mu^M(t,dx) = f(t,x)dx\,,
\end{equation*}
where $f$ is a density, solution of the irreversible LS coagulation model (see details in Annex, section 4), given, for all $t\geq 0$, by
\begin{equation}\label{eq:LS_irr}
\left\lbrace \begin{array}{rcll}
\ds \frac{\partial}{\partial t}f(t,x)+\frac{\partial}{\partial x}(p(x)c_1(t)f(t,x)) & \ds = & \ds 0\,, \quad \forall x>0\,,\\
\ds c_1(t) + \int_0^{\infty}xf(t,x)dx & \ds = & \ds 1\,,\\
\ds \lim_{x\to 0^+} (p(x)f(t,x)) & \ds = & \ds \overline p_1 c_1(t)\,,
 \end{array}
\right.
\end{equation}
with initial condition $c_1(0) = 1$ and $f(0,\cdot) = 0$, and where $p(x)$ is the limit of the macroscopic coagulation rate $p^M(x)$ defined in Eq.~\eqref{eq:macro_rate}. Such Eq.~\eqref{eq:LS_irr} is a transport PDE with ingoing characteristics at $x=0^+$, and is well-defined as a boundary condition at $x=0$ is given. We refer to the paper \cite{DHY} for the choice of the boundary condition (that depends on the scaling used in Eq.~\eqref{eq:rescaled_measure} and the scaling of the reaction rates). The large cluster $C_k(t)$ for $k=\sqrt{M}$ is thus approximated by $f(\sqrt{M}t,1)$, which yields
\begin{equation}\label{eq:result1}
\sqrt{M}T^{\sqrt{M},M}_{\rho,h} \approx_{M\to\infty} \inf\{t\geq 0 :f(t,1)\geq \rho M^{h-1/2}\,\}\,.
\end{equation}
The latter quantity is deterministic, and may be finite or infinite, according to the macroscopic coagulation rate $p$ and the threshold $\rho$.

\subsubsection{$N\to\infty$, ``System-size expansion'' scaling}\label{ssec:Ngd_rescale}

Finally, if the coagulation rates are rescaled with the system size $M$, \textit{i.e.} if we choose a rescaled nucleation rate $p_1=\frac{\overline p_1}{M^2}$, and $p_k=\frac{\overline p_k}{M}$, $k\geq2$, then, using $\gamma=-1/2$, we have shown in \cite{DHY} that the that the measure $\mu^M$ statisfies
\begin{equation*}
 \lim_{M\to \infty} \mu^M(t,dx) = f(t,x)dx\,,
\end{equation*}
where $f$ is a density, solution of the LS coagulation-fragmentation model given, for all $t\geq 0$, by
\begin{equation}\label{eq:LS}
\left\lbrace \begin{array}{rcll}
\ds \frac{\partial}{\partial t}f(t,x)+\frac{\partial}{\partial x}\left[(p(x)c_1(t)-q(x))f(t,x)\right] & \ds = & \ds 0\,, \quad \forall x >0\,,\\
\ds c_1(t) + \int_0^{\infty}xf(t,x)dx & \ds = & \ds 1\,,\\
\ds \lim_{x\to 0^+} (x^rf(t,x)) & \ds = & \ds c_1(t)\,,
 \end{array}
\right.
\end{equation}
with initial condition $c_1(0) = 1$ and $f(0,\cdot) = 0$, and where $p(x),q(x)$ are the limit of the macroscopic rate $p^M(x),q^M(x)$ defined in Eq.~\eqref{eq:macro_rate}, and $r\in[0,1[$ is determined through the relation $p(x)\approx_{x\to 0} x^r$. Again, such Eq.~\eqref{eq:LS} is well-defined if a boundary condition at $x=0$ is given when the characteristics are ingoing at $x=0^+$. We refer to \cite{DHY} for the choice of the boundary condition (that depends on the scaling used in Eq.~\eqref{eq:rescaled_measure} and the scaling of the reaction rates). The large cluster $C_k(t)$ for $k=\sqrt{M}$ is now approximated by $f(t/\sqrt{M},1)$, so that
\begin{equation}\label{eq:result2}
\frac{1}{\sqrt{M}}T^{\sqrt{M},M}_{\rho,h} \approx_{M\to\infty} \inf\{t\geq 0 :f(t,1)\geq \rho M^{h-1/2}\,\}\,.
\end{equation}
The latter quantity is deterministic, and may be finite or infinite, according to the macroscopic rates $p,q$ and $\rho$. 

\noindent The results of the last two subsections are illustrated below with the help of numerical simulations. Note that for particular choice of rates $p$ and $q$, one is able to obtain analytically time-dependent solution of Eq.~\eqref{eq:LS_irr} and Eq.~\eqref{eq:LS} (Annex, section 4.1).

\begin{figure}[t]
\centering
\includegraphics{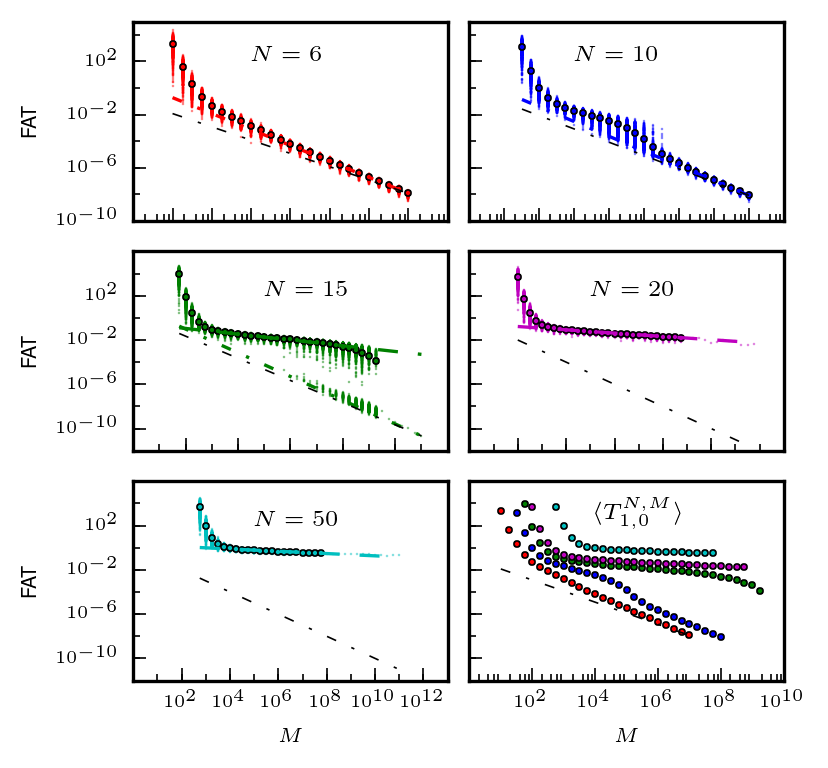}
\caption{First Assembly Time $T^{N,M}_{1,0}$ for the original SBD (section \ref{ssec:noscale}) as a function of the total mass $M$ (in log-log scale) for five different maximal cluster sizes $N\in\{6,10,15,20,50\}$. Each color light dot is a single realization of the FAT. For each condition, large circles represent the statistical mean over $1000$ samples (A few condition are sampled only once, namely for $N=15,20,50$ and large $M$, for which the mean is not shown). Black dash-dotted lines are straight lines of slope $-1$, color dash-dotted lines are straight lines of slope $-(1+1/(N-1))$ (as in Eq.~\eqref{eq:meantime_constant}). And for $N=15,20,50$ we plot additionally dashed lines of slope resp. $-0.26,-0.15$ and $-0.10$. The last panel in bottom-right represent the $5$ mean FAT on the same scale. Kinetic rates are $p_1=0.5$, $p_k\equiv 1$, and $q_k\equiv 100$ for all $k\geq 2$.}
\label{fig:fpt_noscaling_m}
\end{figure}

\begin{figure}[t]
\centering
\includegraphics{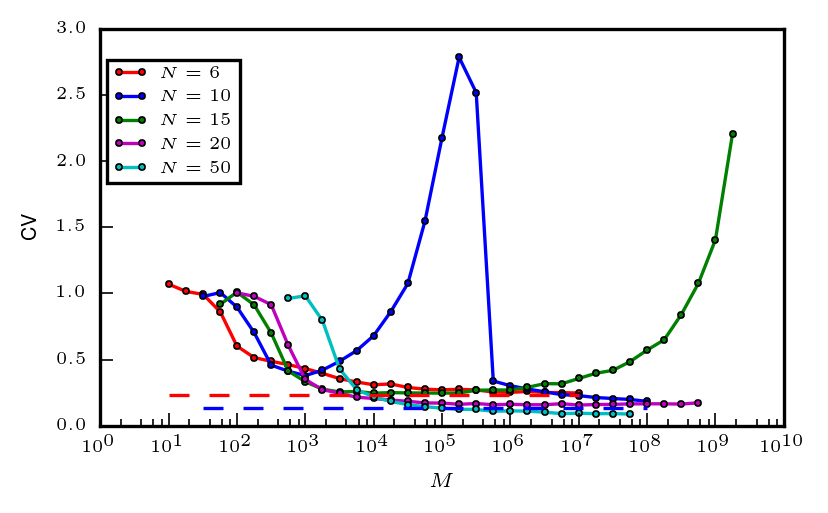}
\caption{Coefficient of Variation (CV) for the First Assembly Time $T^{N,M}_{1,0}$ as a function of the total mass $M$ corresponding to the realizations of Fig~\ref{fig:fpt_noscaling_m}. For $N=6,10$ we plot additionally horizontal dashed lines at the value predicted by the Weibull distribution, see Eq.~\eqref{eq:cvtime_constant}.}
\label{fig:fpt_noscaling_m_cv}
\end{figure}

\section{Simulations and Analysis}

\begin{figure}[t]
\centering
\includegraphics{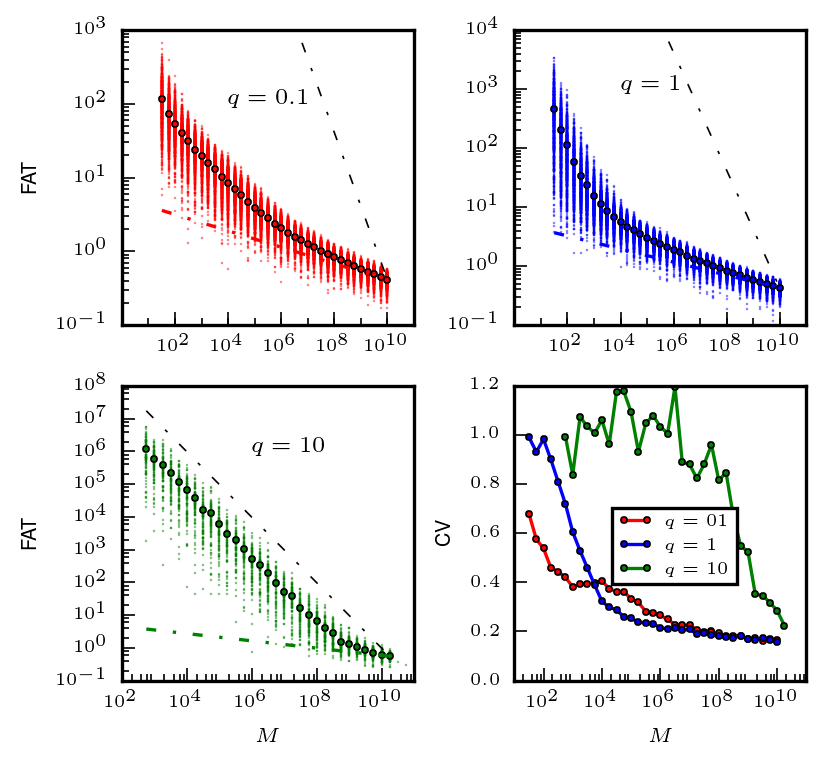}
\caption{First Assembly Time $T^{N,M}_{1,0}$ for the rescaled SBD (section \ref{ssec:system-sizeexpansion}) as a function of the total mass $M$ (in log-log scale) for three different detachment rates $q\in\{0.1,1,10\}$, and $N=10$. Kinetic rates are $p_1=0.5$, and $p_k\equiv 1$ and $q_k\equiv q$  for all $k\geq 2$. Each color light dot is a single realization of the FAT. For each condition, large circles represent the statistical mean over $1000$ samples. Black dash-dotted lines are straight lines of slope $-1$, color dash-dotted lines are straight lines of slope $-1/(N-1)$. Finally, the panel in bottom right represent the Coefficient of Variation (CV) as a function of the total mass $M$ corresponding to the realizations of the first three panels (top and bottom left).}
\label{fig:fpt_scaling_pk}
\end{figure}

\begin{figure}[t]
\centering
\includegraphics{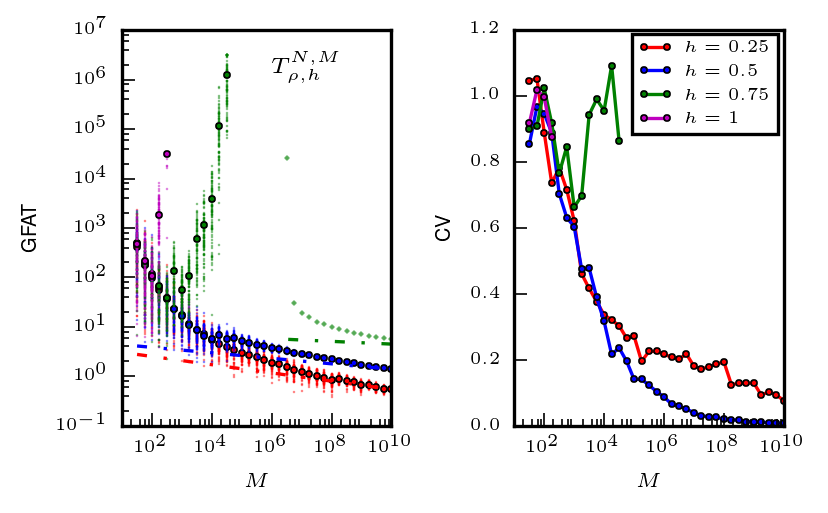}
\caption{(Left) Generalized First Assembly Time $T^{N,M}_{\rho,h}$ for the rescaled SBD (section \ref{ssec:system-sizeexpansion}) as a function of the total mass $M$ (in log-log scale) for $h\in\{0.25,0.5,0.75,1\}$, and $N=10$. Kinetic rates are $p_1=0.5$, $p_k\equiv 1$ and $q_k\equiv 1$ for all $k\geq 2$. Each color light dot is a single realization of the GFAT. For each condition, large circles represent the statistical mean over $1000$ samples (A few condition are sampled only once, namely for $h=0.75$ and large $M$, for which the mean is not shown). Color dash-dotted lines are straight lines of slope $-(1-h)/(N-1)$. (Right) Coefficient of Variation (CV) as a function of the total mass $M$ corresponding to the realizations of the left panel.}
\label{fig:gfpt_scaling_pk}
\end{figure}

\noindent In this section we present results derived from simulations of the SBD model associated to the stochastic Eq.~\eqref{HOMOEQN0}-\eqref{HOMOEQN0_flux}, for various values of its key parameters $\{M,N,p_k,q_k\}$.  Specifically, we use an exact stochastic simulation algorithm (kinetic Monte-Carlo, KMC) to calculate first assembly times \cite{BKL75, GILLESPIE77, GIBSON}. For each set of $\{M,N,p_k,q_k\}$ we sample $10^3$ trajectories (except for few cases where sampling so many trajectories was out of reach in terms of computational time) and follow the time evolution of the cluster populations until the threshold is reached, when the
simulation is stopped and the FAT/GFAT recorded. We compare and contrast our numerical results with the theoretical findings of the previous sections. The following is divided in four sections that corresponds to four main results on the FAT and the GFAT. In all Figures from Fig.~\ref{fig:fpt_noscaling_m} to Fig.~\ref{fig:fpt_largenucl_rescattach_meta} we represent each realization of the FAT (resp. GFAT) in light dot together with its empirical mean in large dot. We superpose on top to it the relevant analytical curves to illustrate the consistency with the theoretical findings.

\subsection{The First Assembly Time can be weakly-dependent on the total number of monomer $M$, and highly variable even in large population}
\noindent We begin with the analysis of the FAT as a function of the total number of monomer $M$, when the maximal cluster size $N$ and the aggregation rates $p_k$ are fixed. For $N=6,10,15$, the prediction of the asymptotic behavior of $T^{N,M}_{1,0}$, the waiting time for a single maximal cluster to be formed, is verified: the mean FAT decrease linearly in log-log scale as $M$ increase, with a slope equal to $-(1+1/(N-1))$, as in the linear CMSBD model (Fig.~\ref{fig:fpt_noscaling_m}, upper panels), see Eq.~\eqref{eq:meantime_constant}. The coefficient of variation (cv, standard deviation over the mean), that measures the variability of the FAT, is also consistent with a transition from an exponential distribution to a Weibull distribution as $M$ increases: the cv decreases from $1$ to the predicted value by Eq.~\eqref{eq:cvtime_constant} (Fig.~\ref{fig:fpt_noscaling_m_cv}, and Fig.~A.1 in Annex for the CMSBD). Furthermore, one can observe very clearly for $N=15$ the bimodal behavior predicted for large but intermediate $M$ values (Fig.~\ref{fig:fpt_noscaling_m}, third panel). For $M$ from $10^6$ to $10^{10}$, the sampled FAT segregates between two groups separated by several order of magnitude (one group below $10^{-6}$, one group around $10^{-2},10^{-3}$). The higher values of the sampled FAT corresponds to trajectories that went through the threshold $C_N=1$ after the metastable period described in paragraph \ref{ssec:noscale}. For $N=20$ and $N=50$, we could simulate in a reasonable computation time (several weeks) only up to $M=10^{13}$ and $M=10^{11}$ respectively. Below these values, the metastable states computed in table \ref{table:pi} predict that the threshold will be mostly reached after the metastable period, which explain the large 'plateau' observed for the FAT up to $M^{13}$ (resp $M^{11})$: the FAT is nearly independent of $M$ on a broad range of values (Fig.~\ref{fig:fpt_noscaling_m}, the slope for $N=15,20,50$ are resp. approximatively $-0.26,-0.15,-0.10$). The bimodal behavior observed for the FAT for $N=10,15$ can also be visualized on the cv, which results in a large peak of the cv values for intermediate $M$ values (Fig.~\ref{fig:fpt_noscaling_m_cv}). Trajectories of the number of cluster as a function of time help to visualize the different phases. We illustrate in Annex, Fig.~A.7,~A.8, stochastic trajectories of the SBD model together with the favorable aggregation limit of the deterministic BD model, in order to clearly identify the metastable period. In Fig.~A.9 and~A.10, we exhibit two trajectories of the stochastic SBD model that results in two FAT that differ from several log of order of magnitude, due to the metastable period. We also point the accuracy of the approximation by a linear model that has for initial condition the metastable state. Finally, the transition from an exponential distribution to a Weibull distribution as $M$ increases, trough an intermediate bimodal distribution, is also illustrated on Histogramms of the FAT over $10^3$ realization in Fig.~A.11.

\noindent Similar results are obtained for the GFAT $T^{N,M}_{\rho,h}$, where the linear log-dependence with a slope $-(1+(1-h)/(N-1))$ (see Eq.~\eqref{eq:meantime_constant_GFAT} for the CMSBD model) is found to be perfectly satisfied for $N=3,5$ and $h=0.25,0.5,0.75$ and $h=1$ (Annex, Fig.~A.5, upper panels). Bimodal behavior and nearly flat log-dependence of the GFAT $T^{N,M}_{\rho,h}$ as a function of $M$ on a broad range of $M$ values is also observed for $N=10,20$ (Annex, Fig.~A.5, lower panels). The size of the 'bimodal' region is found to be increased with increasing $h$ (and $N$). For $N=10,20$ and $h=1$, the mean FAT is increasing to $\infty$ as the deterministic limit given by Eq.~\eqref{eq:gfat_largeM_noscaling} is infinite. The cv are non-monotonic with respect to $M$ with a peak corresponding to the bimodal behavior (Annex, Fig.~A.6). We show that the GFAT has a lower variability as $h$ increase, and vanish for $h=1$ and large $M$, in agreement with the deterministic limit in Eq.~\eqref{eq:gfat_largeM_noscaling}.

\subsection{The First Assembly Time is non-monotonic with respect to the detachment rate}

\noindent We verify in Annex, Fig.~A.12,~A.13 and~A.14 the dependence of the FAT on the detachment rate (see also\cite{YOC}). We confirm that the bimodal behavior is observed for \textit{small} detachment rate, and that the mean FAT (and the cv) is a non-monotonic function of the detachment rate.

\begin{figure}[t]
\centering
\includegraphics{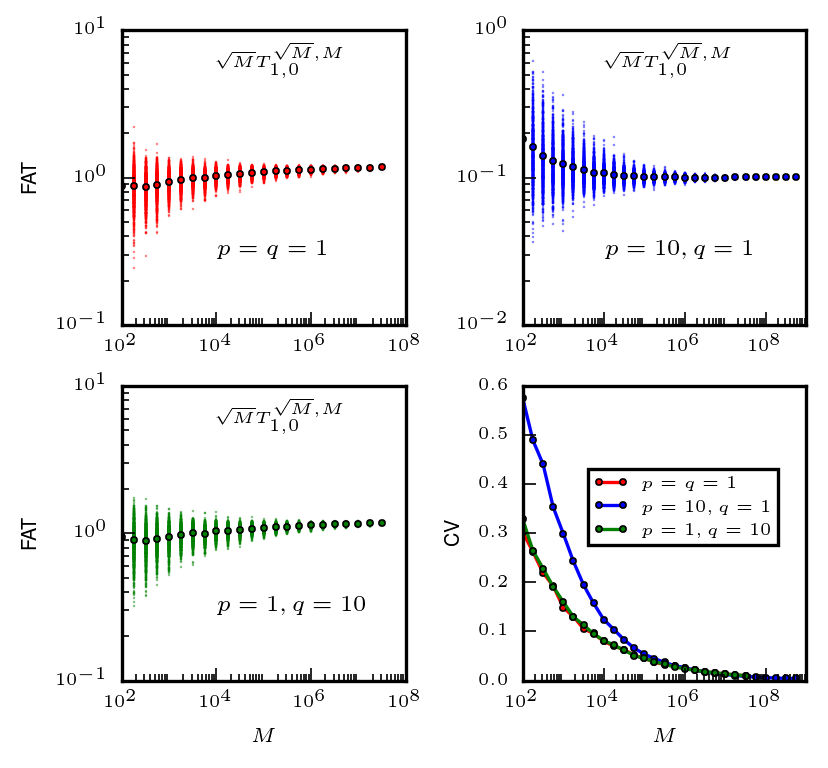}
\caption{First Assembly Time $T^{\sqrt{M},M}_{1,0}$ for the original SBD and large maximal cluster size of order $N=\sqrt{M}$ (section \ref{ssec:Ngd_norescale}) as a function of the total mass $M$ (in log-log scale) for three different kinetic rates $(p_k,q_k)\in\{(1,1);(10,1);(1,10)\}$. The FAT is multiplied by $\sqrt{M}$ to verify the scaling in Eq.~\eqref{eq:result1}. Finally, the panel in bottom right represent the Coefficient of Variation (CV) as a function of the total mass $M$ corresponding to the realizations of the first three panels (top and bottom left).}
\label{fig:fpt_largenucl_rescnucl}
\end{figure}

\subsection{The Generalized First Assembly Time may increase with $M$ for the system-size scaling}

\noindent When the aggregation rates $p_k$ are rescaled with the total number of monomer $M$ (see section \ref{ssec:system-sizeexpansion}), the FAT to reach a maximal cluster of fixed size $N$ decrease monotonically with $M$, and asymptotically with a linear log-dependence with a slope $1/(N-1)$ (Fig.~\ref{fig:fpt_scaling_pk}), as predicted by Eq.~\eqref{eq:fpt_scaling_pk}. The same is valid for the GFAT $T^{N,M}_{\rho,h}$, for $h<1$, with a slope $(1-h)/(N-1)$ (Fig.~\ref{fig:gfpt_scaling_pk}). However, for $h=1$, if the threshold $\rho$ is too large, the GFAT is never reached by the deterministic BD model~\eqref{eq:BD_det_aggirr}-\eqref{eq:detBD_limit_flux}. Thus, for the finite SBD, the GFAT for $h=1$ increases to $\infty$ as $M$ increases to $\infty$. For $h=0.75$, we also found that the GFAT is \textit{non-monotonic} with respect to the total number of monomers, eventhough it converges to $0$ for (very) large number of monomers.

\begin{figure}[t]
\centering
\includegraphics{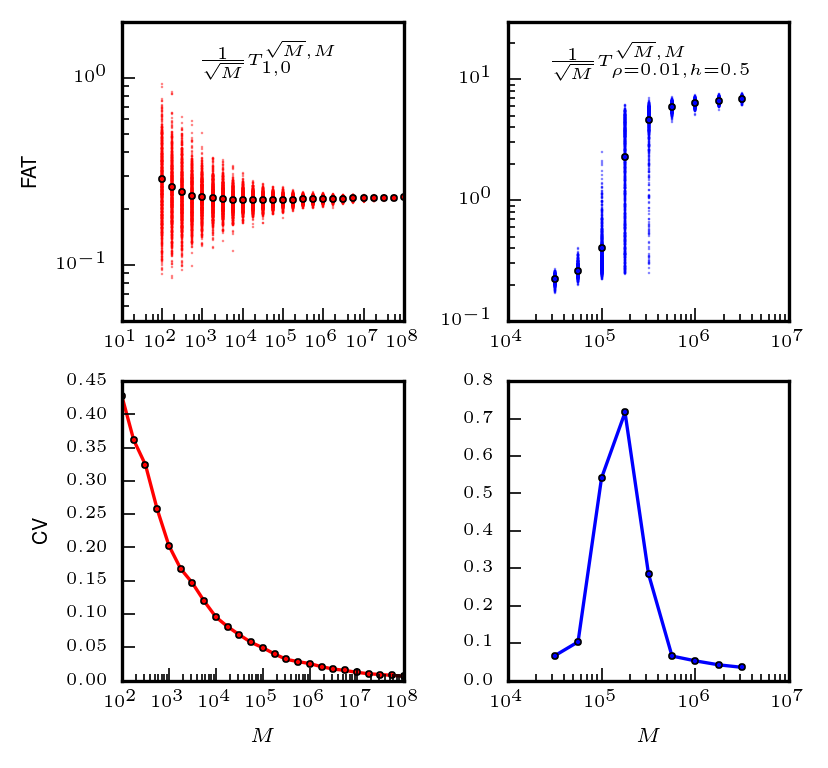}
\caption{First Assembly Time $T^{\sqrt{M},M}_{1,0}$ (top left) and Generalized First Assembly Time $T^{\sqrt{M},M}_{\rho,h}$ (top right) for the rescaled SBD and large maximal cluster size of order $N=\sqrt{M}$ (section \ref{ssec:Ngd_rescale}) as a function of the total mass $M$ (in log-log scale). Kinetics rates are $p(x)\equiv 5$ and $q(x)=x$. Both the FAT and the GFAT are divided by $\sqrt{M}$ to verify the scaling in Eq.~\eqref{eq:result2}. Finally, the panels in bottom left and right represent the Coefficient of Variation (CV) as a function of the total mass $M$ corresponding to the realizations of the upper panels.}
\label{fig:fpt_largenucl_rescattach}
\end{figure}

\begin{figure}[t]
\centering
\includegraphics{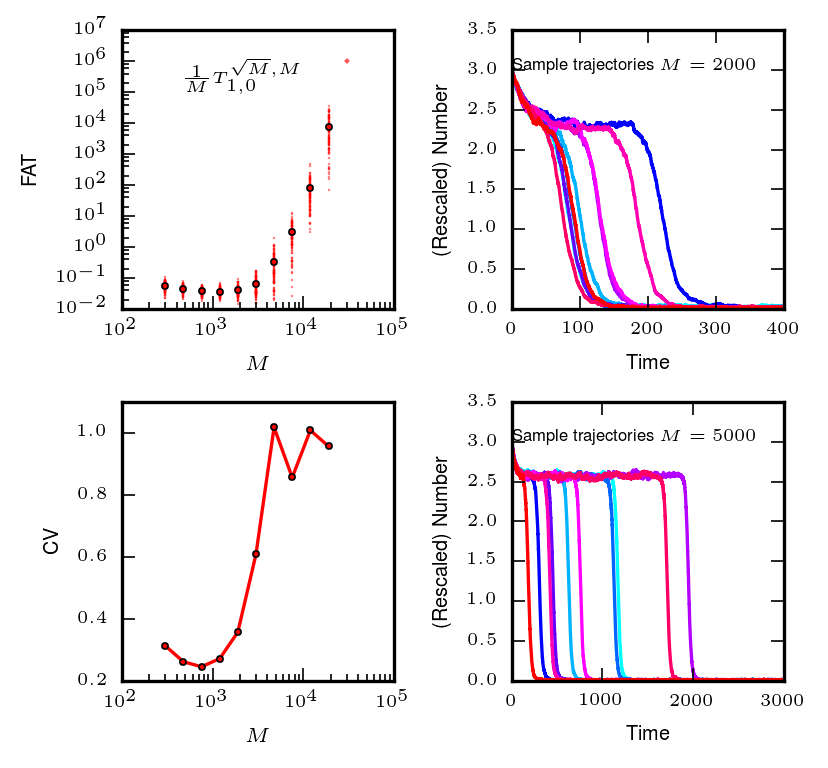}
\caption{(Top left) First Assembly Time $T^{\sqrt{M},M}_{1,0}$ for the rescaled SBD and large maximal cluster size of order $N=\sqrt{M}$ (section \ref{ssec:Ngd_rescale}) as a function of the total mass $M$ (in log-log scale). Kinetics rates are $p(x)=x$ and $q(x)=1$. (Bottom Left) Coefficient of Variation (CV) as a function of the total mass $M$ corresponding to the realizations of the upper left panel. (Top and Bottom Right) Time-dependent trajectories of the rescaled number of monomers $c_1(t)=C_1(t)/M$, for $M=2000$ (top) and $M=5000$ (down). Each color line represent a single realization with the same initial condition and kinetic parameter.}
\label{fig:fpt_largenucl_rescattach_meta}
\end{figure}

\subsection{Exponentially Large FAT for large maximal cluster size $N$ and phase-transition phenomena}

\noindent Finally, we verify in Fig.~\ref{fig:fpt_largenucl_rescnucl} and Fig.~\ref{fig:fpt_largenucl_rescattach}, that for large maximal size $N$, of order $\sqrt{M}$, the two scalings show in Eq.~\eqref{eq:result1}-\eqref{eq:result2} are valid. Specifically, in Fig.~\ref{fig:fpt_largenucl_rescnucl}, we see that for $M>10^{6}$, the FAT is nearly deterministic and can then be predicted by the limit model Eq.~\eqref{eq:LS_irr}. The same threshold is empirically observed in Fig.~\ref{fig:fpt_largenucl_rescattach} for the GFAT as well. However, considering $p(x)=x$ and $q(x)=1$, in Fig.~\ref{fig:fpt_largenucl_rescattach_meta}, we show that exponential large deviation may occur if the aggregation is not favorable compared to the fragmentation, as in the SCSBD model~\eqref{eq:reaction_oneagg} (Annex, Fig.~A.4). Indeed, in such case, the deterministic limit~\eqref{eq:LS} predicts that the FAT is never reached (and equals $\infty$) as the drift is negative for small (macroscopic) size $x$. For the finite system, the FAT grows exponentially fast with $M$, in agreement with Eq.~\eqref{eq:sol_oneagg_Mtoinfty}. On the right panels in Fig.~\ref{fig:fpt_largenucl_rescattach_meta}, we show few time-dependent trajectories that are representative of a phase-transition phenomena, with a very abrupt change of phase, occurring at a widely variable time (the cv is near $1$). Although the deterministic limit predicts that the aggregation will \textit{not} take place (and the monomer number will not decrease), in the SBD model the aggregation is \textit{always} complete (no monomer at the end), but at larger and larger time as $M$ is increasing.

\section{Summary and Conclusions}


We have studied the problem of determining the First Assembly Time (FAT) of a cluster of a pre-determined size $N$ to form from an initial pool of $M$ independent monomers characterized by size-dependent attachment and detachment rates $p_k$ and $q_k$, respectively. We have developed a full stochastic approach, based on the stochastic Becker D\"oring equations (SBD).

\noindent We developed two simplified model and were able to find exact results for the FAT statistics for general values of $M$,$N$ and $p_k$,$q_k$.
The first simplification is to consider that the number of monomer stays constant over time (linear CMSBD model). The FAT was found to be asymptotically (for large $M$) a Weibull distribution, and the mean GFAT decrease to $0$ as $M$ increase to infinity with a log-linear dependence, with coefficient $1+(1-h)/(N-1)$, for any $h\in[0,1]$, and any $N$.
The second simplification is to consider that a single cluster can be formed at a time (SCSBD). The FAT was found to be asymptotically an exponential distribution (for large detachment rate $q_k$), and the mean FAT increase to $\infty$ as $q$ increase to infinity with a log-linear dependence, with coefficient $N-2$, for any $N$. We also show that in the case of unfavorable aggregation ($q_k>p_k$), the formation of large cluster takes an exponentially large time as $M$ increases to $\infty$. 

\noindent With the analytical results on the simplified model in mind, we analyzed the behavior of the FAT for the full SBD. Using a rescaling strategy, as the total number of monomer $M$ increases to $\infty$, we found asymptotic expression of the mean FAT and GFAT as a function of a first passage time associated to deterministic models, namely the discrete-size Becker-D\"oring (BD) model and the continuous-size Lifshitz-Slyozov (LS) model. This has for first implication to be able to find very quickly the order of magnitude of the FAT (resp. GFAT) with the help of a single (fast) numerical simulation of a deterministic model (rather than by extensive numerical simulation of the full SBD model). With the help of a careful time scale analysis on the deterministic BD model, and with extensive numerical simulation, we also pointed out surprising deviations from the mean field deterministic model. First, for sufficiently large maximal cluster size ($N \geq 15$), the mean FAT is found to be very weakly-dependent on the total number of monomers $M$, and so for several order of magnitude of ``intermediate values'' of $M$ (from $10^6$ to $10^{13}$ in our simulations). The full distribution of the FAT is bimodal on this parameter region. We explained and gave practical criteria to observe this phenomena by a careful inspection of the metastable state of the favorable aggregation limit for the deterministic BD model. Second, for large maximal cluster size, we confirmed that for unfavorable aggregation kinetic ($q(x)>p(x)$), the mean FAT is exponentially large as $M$ increases to $\infty$ for the full SBD model. We linked this behavior with phase-transition phenomena, where the number of monomers drastically drops to $0$ in a very short time, compared to the FAT. This phase-transition phenomena occurs as a large deviation from the deterministic model, which predicts that the number of monomers stays constant (no aggregation takes place).

\noindent This study has generalized previous study on the first passage time on the stochastic Becker-D\"oring model. Up to our knowledge, this study is the first one to capture the behavior of the FAT and its generalization for arbitrary kinetic rates, and to explore systematically its dependence with respect to the total number of monomers and the size of the maximal cluster. Taking into account size-dependent kinetic rate is important in practice, as monomer binding and unbinding usually depends on the available surface area of the cluster (for the spherical shape, $p_k\sim k^{2/3}$). This study may have several important applications. One of this is the explanation of the nucleation time observed in \textit{in vitro} polymerization assay of misfolded proteins linked to neurodegenerative diseases\cite{powers06,Xue08,Hingant,YVINEC,Knowles10}. Typical experiments performed in this field are able to record the nucleation time (defined as the time for which the polymerization starts) for various initial quantity of proteins. Some experiments have described a very weak dependence with respect to this initial quantity, where traditional nucleation theory could not explain this fact. Our stochastic approach points out several new behavior that may explain the observations. Furthermore, we argue that having a model that is able to take into account the observed variability on the nucleation time will be important for parameter inference from experimental data (see also the recent preprint\cite{Eugene15}). Indeed, even though the mean FAT may be weakly dependent on the maximal cluster size $N$ (consider the slope of $1+1/(N-1)$ for large $M$), having the observation of the full distribution will facilitate the inference of the maximal cluster size (the shape parameter of the Weibull distribution is $k=N-1$). Finally, on a more theoretical side, the phase-transition phenomena of the SBD model for unfavorable aggregation and large cluster size seems to be described here for the first time. This gives a possible different definition of the nucleation rate, as an inherent infrequent stochastic process, in contrast to classical nucleation theory. It remains in the future to make a link with studies on gelation phenomena, which happen when a fraction of the mass is concentrated in a giant particle ($N$ is of order of $M$). Such studies have been performed mostly in general smoluchowski coagulation models\cite{Jeon97,Fournier09,rezakhanlou}.

\noindent A number of generalization of this model could be considered and will be relevant to tackle new biophysical problems. One could generalize this study to allow general coagulation-fragmentation between any two clusters \cite{ORSOGNA15}. This extension as well as the treatment of heterogeneous nucleation and secondary pathways will be considered in future work.

\section{Acknowledgements}
This work has been supported by ANR grant MADCOW no. ANR-08-JCJC-0135-01 (France), and Association France-Alzheimer, SM 2014.
EH was supported by CAPES/IMPA.

\section{Tables}

 \begin{table}[!ht]
\caption{\textbf{Summary of the First Assembly Time (FAT) and Generalized First Assembly Time (GFAT) findings in this paper. Analytical and numerical results.}}
\centering
\begin{tabular}{|c|c|c|c|}
  \hline
  \rule[-1ex]{0.0cm}{4ex} Model& Condition & $M$ (log-log) dep. & Distrib. \\
  \hline
\rule[-1ex]{0.0cm}{4ex} CMSBD & $M\to\infty$ & $-(1+(1-h)/(N-1))$ & Weibull \\ \hline
\rule[-1ex]{0.0cm}{4ex} SCSBD & $q\to\infty$ & $-N$ & Exponential \\ \hline
\rule[-1ex]{0.0cm}{4ex} SCSBD & $N=\sqrt{M}\to\infty$, $q_k>p_k$ & $Me^{\sqrt{M}}$ & Expo. Large deviation \\ \hline
\rule[-1ex]{0.0cm}{4ex} SBD & $M\to\infty$ & $-(1+(1-h)/(N-1))$ & Weibull \\ \hline
\rule[-1ex]{0.0cm}{4ex} SBD & $Md_N^*\ll 1$ & $\sim 0$ & Bimodal\\ \hline
\rule[-1ex]{0.0cm}{4ex} SBD, $p_k=\overline p_k/M$ & $M\to\infty$ & $-(1-h)/(N-1)$ & \\ \hline
\rule[-1ex]{0.0cm}{4ex} SBD & $N=\sqrt{M}\to\infty$ & $-1/2$ & \\ \hline
\rule[-1ex]{0.0cm}{4ex}  SBD, $p_k=\overline p_k/M$ & $N=\sqrt{M}\to\infty$ & $1/2$ & \\ \hline
\rule[-1ex]{0.0cm}{4ex}  SBD, $p_k=\overline p_k/M$ & $N=\sqrt{M}\to\infty$, $q_k>p_k$ & $Me^{\sqrt{M}}$ & Expo. Large deviation \\ \hline
\end{tabular}
\begin{flushleft}
In this table, we sum up the different analytical findings on the FAT, for the full Stochastic Becker-D\"oring (SBD), Eq.~\eqref{HOMOEQN0}-\eqref{HOMOEQN0_flux} and its two simplifications with constant monomer (CMSBD), Eq.~\eqref{eq:chemreact_constant} and single cluster (SCSBD), Eq.~\eqref{eq:reaction_oneagg}. The first column denotes which model is considered, with which scaling. The second column provides in which asymptotic the results are valid. The third column gives the slope of the log-log dependence of the GFAT with respect to $M$ (except for the SCSBD and SBD with $N=\sqrt{M}$ where exponential large deviation occurs). The last column gives the full distribution of the FAT (if known). See text for more details.
\end{flushleft}
\label{table:nucleationtime}
\end{table}
\pagebreak
\begin{table}[!ht]
\caption{\textbf{Normalized metastable values $d_k^*$ for the  deterministic BD model~\eqref{eq:BD_det_aggirr}-\eqref{eq:detBD_favoragg_flux} in the favorable aggregation case $pM\gg q$.}}
\centering
\begin{tabular}{|c|c|c|c|}
  \hline
  \rule[-1ex]{0.0cm}{4ex} size & value & size & value \\
  \hline
\rule[-1ex]{0.0cm}{4ex}  $d_2^*$ & $0.1839$ & $d_{10}^*$ & $9.1240 10^{-7}$ \\ \hline
\rule[-1ex]{0.0cm}{4ex} $d_3^*$ & $0.1226$ & $d_{11}^*$ & $9.2162 10^{-8}$ \\ \hline
\rule[-1ex]{0.0cm}{4ex} $d_4^*$ & $0.0460$ & $d_{12}^*$ & $8.4481 10^{-9}$ \\ \hline
\rule[-1ex]{0.0cm}{4ex} $d_5^*$ & $0.0123$ & $d_{13}^*$ & $7.0894 10^{-10}$ \\ \hline
\rule[-1ex]{0.0cm}{4ex} $d_6^*$ & $0.0026$ & $d_{14}^*$ & $5.4858 10^{-11}$ \\ \hline
\rule[-1ex]{0.0cm}{4ex} $d_7^*$ & $4.3795 10^{-4}$ & $d_{15}^*$ &   $3.9385 10^{-12}$ \\ \hline
\rule[-1ex]{0.0cm}{4ex} $d_8^*$ & $6.3868 10^{-5}$ & $d_{20}^*$ &   $2.8730 10^{-18}$ \\ \hline
\rule[-1ex]{0.0cm}{4ex} $d_9^*$ & $8.1102 10^{-6}$ & $d_{50}^*$ &   $5.9269 10^{-64}$ \\ \hline
\end{tabular}
\begin{flushleft}
In this table, we compute the numerical values of the normalized metastable values $d_i^*$ for the deterministic BD model~\eqref{eq:BD_det_aggirr}-\eqref{eq:detBD_favoragg_flux} with constant kinetic rate $p_k\equiv p$ and $q_k\equiv q$ in the favorable case $pM\gg q$. Such values represent the level that each variable reach during the metastable period after the pure-aggregation period. It is given by the equilibrium value of the irreversible BD model~~\eqref{eq:BD_det_aggirr}-\eqref{eq:BD_det_aggirr_flux}, see Eq.~\eqref{eq:meta_BDirr}. See text in subsection ~\ref{ssec:noscale}
\end{flushleft}
\label{table:pi}
\end{table}

\pagebreak

\end{document}